# High-fidelity spin and optical control of single silicon vacancy centres in silicon carbide

**Authors:**

Roland Nagy[1], Matthias Niethammer[1], Matthias Widmann[1], Yu-Chen Chen[1], Péter Udvarhelyi[2,3], Cristian Bonato[4], Jawad Ul Hassan[5], Robin Karhu[5], Ivan G. Ivanov[5], Nguyen Tien Son[5], Jeronimo R. Maze[6,7], Takeshi Ohshima[8], Öney O. Soykal[9], Ádám Gali[2,10], Sang-Yun Lee[11,#], Florian Kaiser[1,*], and Jörg Wrachtrup[1]

# sangyun.lee236@gmail.com

* f.kaiser@pi3.uni-stuttgart.de

**Affiliations:**

1 3rd Institute of Physics, University of Stuttgart and Institute for Quantum Science and Technology IQST, Germany

2 Institute for Solid State Physics and Optics, Wigner Research Centre for Physics, Hungarian Academy of Sciences, Budapest, PO Box 49, H-1525, Hungary

3 Department of Biological Physics, Eötvös Loránd University, Pázmány Péter sétány 1/A, H-1117 Budapest, Hungary

4 Institute of Photonics and Quantum Sciences, SUPA, Heriot-Watt University, Edinburgh EH14 4AS, United Kingdom

5 Department of Physics, Chemistry and Biology, Linköping University, SE-58183 Linköping, Sweden

6 Facultad de Física, Pontificia Universidad Católica de Chile, Santiago 7820436, Chile

7 Research Center for Nanotechnology and Advanced Materials CIEN-UC, Pontificia Universidad Católica de Chile, Santiago 7820436, Chile

8 National Institutes for Quantum and Radiological Science and Technology, Takasaki, Gunma 370-1292, Japan

9 Naval Research Laboratory, Washington, D.C. 20375, USA

10 Department of Atomic Physics, Budapest University of Technology and Economics, Budafoki út 8, H-1111 Budapest, Hungary

11 Center for Quantum Information, Korea Institute of Science and Technology, Seoul, 02792, Republic of Korea

**Abstract paragraph:**

Optically interfaced spins in the solid promise scalable quantum networks. Robust and reliable optical properties have so far been restricted to systems with inversion symmetry. Here, we release this stringent constraint by demonstrating outstanding optical and spin properties of single silicon vacancy centres in silicon carbide. Despite the lack of inversion symmetry, the system's particular wave function symmetry decouples its optical properties from magnetic and electric fields, as well as from local strain. This provides a high-fidelity spin-to-photon interface with exceptionally stable and narrow optical transitions, low inhomogeneous broadening, and a large fraction of resonantly emitted photons. Further, the weak spin-phonon coupling results in electron spin coherence times comparable with nitrogen-vacancy centres in diamond. This allows us to demonstrate coherent hyperfine coupling to single nuclear spins, which can be exploited as qubit memories. Our findings promise quantum network applications using integrated semiconductor-based spin-to-photon interfaces.

**Main text:**

Optically addressable single spins in solids are a promising basis for establishing a scalable quantum information platform[1–3]. Electron spins are naturally suited for nanoscale quantum sensing[4–9], and for controlling nuclear spins that enable quantum information storage and computation[10–12]. Combined with an efficient spin-to-photon interface[13,14], this enables fast optical spin manipulation[13,15], entangling multiple quantum systems over long distances[16,17], and the realisation of quantum networks[1,18]. In this perspective, several pivotal landmark demonstrations have been achieved with the nitrogen-vacancy defect centre in diamond[13,16,18,19]. However, implementation of nano-photonics structures in the vicinity of spin-based quantum systems usually severely compromises spin and optical stability and coherence[20,21]. Therefore, new systems are needed that are decoupled from detrimental interactions by a high degree of symmetry. For example, the negatively charged silicon vacancy defect in diamond shows very high optical stability due to inversion symmetry at low strain[22,23]. However, the pronounced spin-phonon coupling necessitates millikelvin temperatures for practical spin coherence times[24]. Alternatively, one can resort to systems showing highly symmetric ground and excited state symmetries and little redistribution of electron density between the two states, but with alternating phase of the corresponding wavefunctions that maximises the transition dipole moment. Such systems show strong zero phonon lines and potentially little changes in electric dipole moment between ground and excited state, *i.e.* small coupling to electric field fluctuations. So far, no known quantum system in solids demonstrated all the above specifications simultaneously. Here we show that the silicon vacancy ($V_{Si}$) in the 4H polytype of silicon carbide (SiC) close to ideally matches all criteria. By demonstrating that inversion symmetry is not a stringent criterion for an ideal quantum emitter, we open the door for a new class of quantum systems in numerous semiconductors and insulators.

The structure of the 4H-SiC crystal results in two non-equivalent sites for a $V_{Si}$. As shown in Figure 1(a), we investigate the defect centre that is formed by a missing silicon atom at a hexagonal lattice site (V1 centre)[25]. The ground state has a spin quartet manifold ($S$=3/2) with weak spin-orbit coupling, leading to milliscond spin relaxation times[26]. To study the centre's intrinsic optical and electron spin properties, we use a nuclear spin free isotopically purified 4H-SiC layer ($^{28}$Si > 99.85%, $^{12}$C > 99.98%), in which we create single centres by electron irradiation (see Methods). The defects are optically addressed by confocal microscopy at cryogenic temperatures (T ~ 4 K). We employ a 730 nm laser diode for off-resonant excitation. A wavelength-tunable diode laser (Toptica DL pro) performs resonant excitation at 861 nm in the zero phonon line of the lowest optical transition with $A_2$ symmetry, known as V1 line. Another energetically higher transition to an electronic state of E symmetry, called V1' line[25–27], is not investigated in this work. Fluorescence emission is detected in the red-shifted phonon side band (875

– 890 nm). In addition, ground state spin manipulation is performed with microwaves (MW) that are applied via a 20 µm diameter copper wire.

Figure 1(b) illustrates the defect's energy level structure. At zero external magnetic field ($B_0 = 0$ G), ground and excited state manifolds show pairwise degenerate spin levels $m_S = \pm 1/2$ and $m_S = \pm 3/2$, with zero-field splittings (ZFS) of $2 \cdot D_{gs}$ and $2 \cdot D_{es}$, respectively. Previous studies constrained $2 \cdot D_{gs} < 10$ MHz[25,26,28] and here we determine $2 \cdot D_{gs} = 4.5 \pm 0.1$ MHz (see Methods). In order to investigate the excited state structure, we use resonant optical excitation. We apply a strong MW field at ~ 4.5 MHz to continuously mix the ground state spin population, and wavelength-tune simultaneously the 861 nm laser across the optically allowed transitions. As shown in Figure 1(c), we observe two strong fluorescence peaks, labelled $A_1$ and $A_2$. The peak separation of $980 \pm 10$ MHz corresponds to the difference between the ground and excited state ZFS. As shown in the supplementary information, we use coherent spin manipulation to infer a positive excited state ZFS, *i.e.* $2 \cdot D_{es} = 985 \pm 10$ MHz, which is in line with the results of first principles density functional theory. To determine optical selection rules, we apply an external magnetic field of $B_0 = 92$ G precisely aligned along the uniaxial symmetry axis of $D_{gs}$, which is parallel to the c-axis of 4H-SiC, such that the ground state Zeeman spin splitting exceeds the optical excitation linewidth (see Figure 1(b)). We observe no shift of the optical resonance lines, corroborating our assignment of $A_{1,2}$ as spin-conserving optical transitions between ground and excited states. Further, we observe no spin-flip transitions, which would show up as additional peaks in the spectra in Fig. 1(c) at approximately $\pm 258$ MHz (Supplementary Information). However, as we will show later, spin-flips can still occur through nonradiative decay channels. Moreover, as the presence of a magnetic field does not alter the peak separation, we confirm that ground and excited state g-factors are identical (Supplementary Information)[29]. Figure 1(d) shows repetitively recorded excitation spectra for which we find exceptionally stable lines over an hour time scale. To underline that the defect's intrinsic symmetry indeed decouples it from strain and stray charges in its local environment, we perform resonant excitation studies on four other defects. As shown in Figure 1(f), the peak separation of all defects is nearly identical. In addition, all resonant absorption lines are inhomogeneously distributed over only a few 100 MHz, allowing us to identify several defects with overlapping emission. Our results suggest a very low strain sensitivity, consistent with the Kramer's degeneracy of half-integer spin systems, and almost identical dipole moment in the ground and excited state (see Supplementary Information). Our findings are intimately related to the symmetry properties of the defect, showing A-type states in ground and first excited state with similar electron density distributions. Unlike inversion symmetry, this does not preclude the existence of an electric dipole moment in ground and excited state but restricts its orientation to the symmetry axis of the defect. A similar line of arguments shows that optical transitions among those states are allowed because of the alternating phase of the wavefunctions nevertheless, if excitation is polarised along the symmetry axis of the defect, *i.e.* the c-axis of the crystal. This is indeed found in our experiments.

Advanced quantum information applications based on spin-photon entanglement require that the quantum system emits transform-limited photons. We measure this via the excitation linewidth of the optical transitions. In Figure 1(e), we show that for excitation intensities below 1 W/cm², the linewidth approaches 60 MHz. Considering the 5.5 ns excited state lifetime[26], this is only twice the Fourier-transform limit, and might be explained by small residual spectral diffusion.

Realising a spin-to-photon interface for quantum information applications requires high-fidelity spin state initialisation, manipulation and readout[30]. Previous theoretical models[31] and ensemble-based measurements[26] have indicated that continuous off-resonant optical excitation of $V_{Si}$ eventually leads to a decay into a metastable state manifold (MS), followed by rather non-selective relaxation into the ground state spin manifold. Here, we use resonant optical excitation to strongly improve spin state selectivity. As shown in Figure 2(a), we apply a magnetic field of $B_0 = 92$ G, allowing us to selectively address transitions within the ground state spin manifold via MW excitation. We first excite the system along the $A_2$ transition (linking the $m_S = \pm 3/2$ spin states), which eventually populates the system into the $m_S = \pm 1/2$ ground state via decays through the MS. Then, we perform optically detected magnetic resonance (ODMR). For this, we apply narrowband microwave pulses in the range of 245 – 275 MHz, followed by fluorescence detection during an optical readout pulse on the $A_2$ transition. As shown in Figure 2(b), we observe two spin resonances at the magnetic dipole allowed transitions, $\left|-\frac{1}{2}\right\rangle_{gs} \leftrightarrow \left|-\frac{3}{2}\right\rangle_{gs}$ and $\left|+\frac{1}{2}\right\rangle_{gs} \leftrightarrow \left|+\frac{3}{2}\right\rangle_{gs}$ at 253.5 MHz (MW$_1$) and 262.5 MHz (MW$_3$), respectively. To observe the centre resonance at 258.0 MHz (MW$_2$), we need to imbalance the population between $\left|-\frac{1}{2}\right\rangle_{gs}$ and $\left|+\frac{1}{2}\right\rangle_{gs}$. To this end, we combine the initial resonant laser excitation pulse on $A_2$ with MW$_3$. MW$_3$ continuously depopulates $\left|+\frac{1}{2}\right\rangle_{gs}$, such that the system is eventually pumped into $\left|-\frac{1}{2}\right\rangle_{gs}$. After applying microwaves in the range between 245 – 275 MHz, the optical readout pulse is also accompanied by MW$_3$, which transfers population from $\left|+\frac{1}{2}\right\rangle_{gs}$ to $\left|+\frac{3}{2}\right\rangle_{gs}$. This effectively allows us to detect a fluorescence signal from spin population in $\left|+\frac{1}{2}\right\rangle_{gs}$. As shown in Figure 2(c), we observe the two expected resonances for $\left|-\frac{1}{2}\right\rangle_{gs} \leftrightarrow \left|-\frac{3}{2}\right\rangle_{gs}$ and $\left|-\frac{1}{2}\right\rangle_{gs} \leftrightarrow \left|+\frac{1}{2}\right\rangle_{gs}$ at 253.5 MHz (MW$_1$) and 258.0 MHz (MW$_2$), respectively. Unless mentioned otherwise, we now always initialise the system into $\left|-\frac{1}{2}\right\rangle_{gs}$ using the above-described procedure and read out the spin state population of the $\left|\pm\frac{3}{2}\right\rangle_{gs}$ levels with a 1 μs long laser pulse resonant with the $A_2$ transition.

We first demonstrate coherent spin state control via Rabi oscillations. Figure 3(a) shows the sequence for Rabi oscillations between $\left|-\frac{1}{2}\right\rangle_{gs} \leftrightarrow \left|-\frac{3}{2}\right\rangle_{gs}$. After spin state initialisation, we apply MW$_1$ for a time $\tau_{\text{Rabi}}$ to drive population towards $\left|-\frac{3}{2}\right\rangle_{gs}$, followed by optical readout. To observe oscillations between $\left|-\frac{1}{2}\right\rangle_{gs} \leftrightarrow \left|+\frac{1}{2}\right\rangle_{gs}$, the system is initialised, MW$_2$ is applied for a time $\tau_{\text{Rabi}}$, followed by a population transfer from $\left|+\frac{1}{2}\right\rangle_{gs} \rightarrow \left|+\frac{3}{2}\right\rangle_{gs}$ using a $\pi$-pulse at MW$_3$, and eventual state readout. Figure 3(b) shows the experimental results from which we deduce Rabi oscillation frequencies of 257.5 kHz and 293.8 kHz. The frequency ratio of $1.14 \approx \sqrt{4/3}$ is in excellent agreement with the theoretical expectation for a quartet spin system[32]. In a next step, we proceed to measuring spin coherence times between the ground state levels $\left|-\frac{1}{2}\right\rangle_{gs}$ and $\left|-\frac{3}{2}\right\rangle_{gs}$. To this end, we perform a free induction decay (FID) measurement in which we replace the Rabi pulse in Figure 3(a) by two $\frac{\pi}{2}$-pulses separated by a waiting time $\tau_{\text{FID}}$. The experimental data in Figure 3(c) show a dephasing time of $T_2^* = 30 \pm 2$ μs, which is comparable with state-of-the-art results reported for NV centres in isotopically ultrapure diamonds[33]. To measure the spin coherence time $T_2$, we use a Hahn-echo sequence by adding a refocusing $\pi$-pulse in the middle of two $\frac{\pi}{2}$-pulses. As shown in Figure 3(d), we measure a spin coherence time of $T_2 = 0.85 \pm 0.12$ ms. For an isotopically purified system, like the one used in the present experiment we expect somewhat longer dephasing times. As outlined in the Supplementary Information, we attribute the origin to be nearby defect clusters created by microscopic cracks induced by the CVD wafer cutting process using a dicing saw. Hence, optimised sample growth, cutting and annealing should increase dephasing times to several tens of milliseconds[34].

Quite interestingly, the initial part of the spin echo decay of this particular defect shows pronounced oscillations resulting from the hyperfine coupling of the electron spin to nuclear spins. Being initially polarized along the *z*-axis in the Bloch sphere, and with microwave pulses polarized along the *x*-axis, the echo sequence measures[35]

$$-\langle S_y \rangle = 1 - \frac{1}{k}[2 - 2\cos(\omega_\alpha \tau) - 2\cos(\omega_\beta \tau) + \cos(\omega_- \tau) + \cos(\omega_+ \tau)]. \quad (1)$$

Here, $k$ is the modulation depth parameter $k = \left(\frac{2\omega_\alpha \omega_\beta}{A_\perp \omega_I}\right)^2$, with $\omega_I$ being the nuclear Larmor frequency, and $\omega_{\alpha,\beta} = \left[(\omega_I + m_{\alpha,\beta} A_\parallel)^2 + (m_{\alpha,\beta} A_\perp)^2\right]^{\frac{1}{2}}$. Here, $m_\alpha = -\frac{3}{2}$ and $m_\beta = -\frac{1}{2}$ are the spin projections of the involved ground states with respect to the z-axis. Further, $\omega_\pm = \omega_\alpha \pm \omega_\beta$, and $A_{\perp,\parallel}$ are the orthogonal and parallel hyperfine components, respectively. Essentially, $\langle S_y \rangle$ is modulated because of quantum beats between hyperfine levels, which not only show the nuclear frequencies $\omega_{\alpha,\beta}$ but also their sum and difference frequencies $\omega_\pm$. The Fourier transformation reveals strong frequency components at $\omega_\alpha/2\pi = 77.9 \pm 0.1$ kHz and $\omega_\beta/2\pi = 76.0 \pm 0.1$ kHz. As $\omega_+$ is quite close to twice the Larmor frequency of a $^{29}$Si nuclear spin ($\omega_I/2\pi \approx 77.9$ kHz), we conclude that the parallel hyperfine coupling $A_\parallel$ is weak compared to the Larmor frequency. In addition, from the inferred modulation depth parameter ($k = 0.15 \pm 0.02$), we infer that there is a sizeable difference between $A_\perp$ and $A_\parallel$. Indeed, from the fit to the data, we infer a purely dipolar coupling with strengths of $A_\perp \approx 29$ kHz and $A_\parallel \approx 10$ kHz, respectively. This allows inferring the relative position of the nuclear spin, using $A_\parallel = \frac{\eta_{Si}}{r^3}(3\cos^2\theta - 1)$ and $A_\perp = \frac{\eta_{Si}}{r^3}(3\sin\theta\cos\theta)$, in which $\eta_{Si} = 15.72$ MHz $\cdot$ Å$^3$ is the dipole-dipole interaction coefficient[36], $r$ is the distance between electron and nuclear spin, and $\theta$ is the c-axis and $r$. We obtain $r \approx 11.6$ Å and $\theta \approx 61°$.

Besides excellent coherence times, a crucial requirement for quantum information applications is high-fidelity quantum state initialisation[30]. To optimise the spin state initialisation procedure, we vary the time interval $\tau_{init}$ of the initialising laser ($A_2$) and microwave fields (MW$_3$) and extract the populations in the four ground states via the contrast of Rabi oscillation measurements (for more details, see Supplementary Information). Figure 4(a) shows the experimental sequence. We first equilibrate all four ground state populations to within < 1% with a 40 μs long off-resonant laser pulse[26]. Then, we apply the selective spin state initialisation procedure for a time $\tau_{init}$, followed by different Rabi sequences, and eventual state readout along the $A_2$ optical transition. Figure 4(b) shows the development of the extracted ground state populations for $\tau_{init}$ ranging from 0 μs to 80 μs. We achieve initialisation fidelities up to $97.5 \pm 2.0\%$, which is comparable to previous demonstrations with colour centres in diamond or SiC[14,37].

The silicon vacancy centre in 4H-SiC satisfies a number of key requirements for an excellent solid-state quantum spintronics system. The high spectral stability and close to transform limited photon emission as well as large Debye Waller factor (>0.4)[26] promise good spin-photon entanglement generation rates. We also mention that almost no defect ionisation has been observed, except at high optical excitation powers beyond saturation. This is in stark contrast to NV centres in diamond where charge state verification is required before each experimental sequence[18]. By removing this requirement, spin-photon entanglement rates are further speed up. The emission wavelength of 861 nm is favourable for fibre based long distance communication as ultralow-noise frequency converters to the telecom range already exist[38,39]. The measured overall emission rates are currently about $20 \cdot 10^3$ counts/s. Given the limited system detection efficiency of around 0.1%, due to the use of non-optimised optics and detectors, we estimate an overall photon emission rate of $\approx 10^7$ s$^{-1}$, limited by non-radiative intersystem crossings. Implementation into optical resonators[40] could increase emission rates by reducing the excited state lifetime, and achieving a Purcell factor of 10-20 would already yield sufficient photon rates to accomplish quantum non-demolition readout of single spin states. The defect's very low strain coupling makes this strategy actually very promising. We showed also that spin coherence times, as well as readout and initialization fidelities are comparable with other systems, e.g. defects in diamond. This allowed us to observe hyperfine coupling, which promises access to long-lived quantum memories and quantum registers. In this perspective, the relatively small ground state ZFS, which is on the order of the typical hyperfine interaction, may prove to be beneficial for nuclear spin polarisation techniques, potentially leading to applications in high-contrast magnetic resonance imaging[41,42]. Additionally, excitation to the second excited state [26,27,31] is predicted to show significantly reduced intersystem crossing rates, thus further enhancing optical spin manipulation capabilities, similar to protocols based on cold atoms or ions. We believe therefore that the silicon vacancy centre in 4H-SiC has a great potential to become a working-horse system in a variety of quantum information applications.

**Methods**

**Sample preparation.**

The starting material is a 4H-$^{28}$Si$^{12}$C silicon carbide layer grown by chemical vapour deposition (CVD) on a n-type (0001) 4H-SiC substrate. The CVD layer is ~ 110 µm thick. The isotope purity is estimated to be $^{28}$Si>99.85% and $^{12}$C>99.98%, which was confirmed by secondary ion mass spectroscopy (SIMS) for one of the wafers in the series. After chemical mechanical polishing (CMP) of the top layer, the substrate was removed by mechanical polishing and the final isotopically enriched free-standing layer had a thickness of ~ 100 µm. Current-voltage measurements at room temperatures show that the layer is n-type with a free carrier concentration of ~ $6 \cdot 10^{13}$ cm$^{-3}$. This value is close to the concentration of shallow nitrogen donors of ~ $3.5 \cdot 10^{13}$ cm$^{-3}$, which was determined by photoluminescence at low temperatures. Deep level transient spectroscopy measurements show that the dominant electron trap in the layer is related to the carbon vacancy with a concentration in the mid $10^{12}$ cm$^{-3}$ range. Minority carrier lifetime mapping of the carrier shows a homogeneous carrier lifetime of ~ 0.6 µs. Since the lifetime was measured by an optical method with high injection, the real lifetime is expected to be double, *i.e.* ~ 1.2 µs.[43] Such a high minority carrier lifetime indicates that the density of all electron traps should not be more than mid $10^{13}$ cm$^{-3}$.[44] To generate a low density of silicon vacancy centres, we used room temperature electron beam irradiation at 2 MeV with a fluence of $10^{12}$ cm$^{-2}$. The irradiation creates also carbon vacancies, interstitials, anti-sites and their associated defects, but their concentrations are expected to be below mid $10^{12}$ cm$^{-3}$. After irradiation, the sample was annealed at 300°C for 30 minutes to remove some interstitial-related defects. In order to improve light extraction efficiency out of this high refractive index material ($n \approx 2.6$), we fabricate solid immersion lenses using a focused ion beam milling machine (Helios NanoLab 650). The sample was cleaned for two hours in peroxymonosulfuric acid to remove surface contaminations[45].

**Experimental setup.**

All the experiments were performed at a cryogenic temperature of 4 K in a Montana Instruments Cryostation. A home-built confocal microscope[26] was used for optical excitation and subsequent fluorescence detection of single silicon vacancies. Off-resonant optical excitation of single silicon vacancy centres was performed with a 730 nm diode laser. For resonant optical excitation at 861.4 nm towards V1 excited state we used an external cavity tunable diode laser (Toptica DL pro). The used laser power for the resonant excitation was adjustable from 0.5 to 500 nW. Fluorescence was collected in the red-shifted phonon sideband (875 – 890 nm) for which we used a tunable long-pass filter (Versa Chrome Edge from Semrock). To detect light, we used a near infrared enhanced single-photon avalanche photodiode (Excelitas). The used 4H-SiC sample was flipped to the side, *i.e.* by 90° compared to the c-axis, such that the polarization of the excitation laser was parallel to the c-axis (E||c) which allows to excite the V1 excited state with maximum efficiency[26,27]. Note that the solid immersion lenses (SILs) were fabricated on this side surface. In order to manipulate ground state spin populations, microwaves are applied through a 20 µm thick copper wire located in close proximity to the investigated V1 defect centres.

**Inferring ground state spin initialisation fidelity.**

Near deterministic ground state spin initialization has been demonstrated via optically pumping assisted by microwave spin manipulation (see Figure 4(a)). We first apply the off-resonant laser excitation (730 nm) to prepare non-initialized spin states. We then initialize the system into $|-\frac{1}{2}\rangle_{\text{gs}}$ by resonant optical excitation along the $A_2$ transition, accompanied by microwaves resonant with the transition $|+\frac{1}{2}\rangle_{gs} \leftrightarrow |+\frac{3}{2}\rangle_{gs}$ (MW$_3$). To determine the population in each ground state spin level, we perform Rabi oscillations for the three allowed transitions, linking the levels $|-\frac{1}{2}\rangle_{gs} \leftrightarrow |-\frac{3}{2}\rangle_{gs}$ (MW$_1$), $|-\frac{1}{2}\rangle_{gs} \leftrightarrow |+\frac{1}{2}\rangle_{gs}$ (MW$_2$), and $|+\frac{1}{2}\rangle_{gs} \leftrightarrow |+\frac{3}{2}\rangle_{gs}$ (MW$_3$). Then, we read out the spin population in $|\pm\frac{3}{2}\rangle_{gs}$ by resonant excitation along the $A_2$ transition for 150 ns. Within such a short readout time, we can safely assume that the obtained fluorescence signal is proportional to the population in $|\pm\frac{3}{2}\rangle_{gs}$. Note that in order to obtain a signal from Rabi oscillations $|-\frac{1}{2}\rangle_{gs} \leftrightarrow |+\frac{1}{2}\rangle_{gs}$, an additional population swap ($\pi$-pulse at MW$_3$) between $|+\frac{1}{2}\rangle_{gs}$ and $|+\frac{3}{2}\rangle_{gs}$ is applied before state readout (see also Figure 3(a)). We denote now the populations in all four ground states by $p_i$ where $i = \{-\frac{3}{2}, -\frac{1}{2}, +\frac{1}{2}, +\frac{3}{2}\}$ stands for the spin quantum number of each state. We then measure the fringe visibility of the obtained Rabi oscillation signal in order to infer ground state spin populations. The fringe visibility for Rabi oscillations between sublevels $|i\rangle_{gs}$ and $|j\rangle_{gs}$ with $i, j = \{-\frac{3}{2}, -\frac{1}{2}, +\frac{1}{2}, +\frac{3}{2}\}$ and $|i - j| = 1$ is defined as,

$$v_{i,j} = \frac{I_{\max} - I_{\min}}{I_{\max} + I_{\min}},$$

where $I_{\max}$ ($I_{\min}$) denotes the maximum (minimum) signal during the Rabi oscillation. Considering that our state readout is only sensitive to spin population in $|\pm\frac{3}{2}\rangle_{gs}$, the three conducted Rabi oscillation experiments lead to the following fringe visibilities:

$$v_{3/2,1/2} = (p_{1/2} - p_{3/2}) / (2 \cdot p_{-3/2} + p_{1/2} + p_{3/2})$$

$$v_{1/2,-1/2} = (p_{-1/2} - p_{1/2}) / (2 \cdot p_{-3/2} + p_{-1/2} + p_{1/2})$$

$$v_{-1/2,-3/2} = (p_{-1/2} - p_{-3/2}) / (2 \cdot p_{3/2} + p_{-3/2} + p_{-1/2}).$$

In addition, the total population of all the ground states must be sum up to unity, *i.e.*

$$p_{-3/2} + p_{-1/2} + p_{+1/2} + p_{+3/2} = 1.$$

As all observed Rabi oscillations start with a local minimum in fluorescence intensity, we can further assume that after initialization, the ground state populations fulfil:

$$p_{-3/2} \sim p_{+3/2} < p_{+1/2} < p_{-1/2}.$$

Solving the system of four equations under these constraints allows us to extract all four values of $p_i$, which is shown for the initialization times 0 µs and 80 µs in Figure 4(b).

**Magnetic field alignment.**

In order to allow for selective ground state spin manipulation, level degeneracy has to be lifted. We do so by applying a magnetic field that has to be precisely aligned along the *z*-axis of the spin system in order to avoid the appearance of new mixed eigenstates. The *z*-axis of the spin system is parallel to the c-axis of the 4H-SiC crystal[28]. We apply a magnetic field of strength $B_0 \sim 92$ G with two permanent magnets, placed outside the cryostat chamber. From previous studies, it is already known that the quantization axis of the silicon vacancy centre on a cubic lattice site (usually referred as V2 centre)[25,45] is parallel to the defect centre studied in this report[28]. Therefore, we take advantage of an ensemble of V2 centres found at the edge of the solid immersion lens, which was likely created by ion bombardment during the FIB milling. For a perfectly aligned magnetic field, the splitting between two outer spin resonances of the V2 centre is $4|D_{gs,V2}| = 140.0$ MHz for $|B_0| > |2 \cdot D_{gs,V2}|$[9,45]. Here, we measure a splitting of $139.93 \pm 0.04$ MHz, meaning that the magnetic field is aligned within $\sim 1.3$ degrees. From ground state spin spectra as in Figures 2(b,c) at the aligned magnetic field, we determine the ZFS of the V1 centre ground state, $2 \cdot D_{gs} = 4.5 \pm 0.1$ MHz.

**Acknowledgements:**

We would like to warmly thank Sébastien Tanzilli, Sina Burk, Rainer Stöhr, Roman Kolesov, Philipp Neumann, Sebastian Zeiser, Matthias Pfender, Thomas Öckinghaus, Charles Babin, Durga Dasari, Ilja Gerhardt and Stephan Hirschmann for technical help and fruitful discussions.

R.N. acknowledges support by the Carl-Zeiss-Stiftung. R.N., M.M., M.W., Y-C.C., F.K. and J.W. acknowledge support by the European Research Council (ERC) grant SMel, the European Commission Marie Curie ETN "QuSCo" (GA No 765267), the Max Planck Society, the Humboldt Foundation, and the German Science Foundation (SPP 1601). S-Y.L. acknowledges support by the KIST institutional program (2E272801), C.B. acknowledges support by the EPSRC (Grant No. EP/P019803/1), and the Royal Society. N.T.S. received support from the Swedish Research Council (VR 2016-04068 and VR 2016-05362), the Carl Trygger Stiftelse för Vetenskaplig Forskning (CTS 15:339), and the Swedish Energy Agency (43611-1). T.O. acknowledges support from JSPS KAKENHI 17H01056; A.G. acknowledges the Hungarian NKFIH grants No. NN118161 of the EU QuantERA Nanospin project as well as the National Quantum Technology Program (Grant No. 2017-1.2.1-NKP-2017-00001). A.G. and J.W. acknowledge the EU-FET Flagship on Quantum Technologies through the project ASTERIQS. F.K. and J.W. acknowledge the EU-FET Flagship on Quantum Technologies through the project QIA.

**Author contributions:**

R.N., S-Y.L., F.K. and J.W. conceived and designed the experiment; R.N., and F.K. performed the experiment; R.N., P.U., C.B., J.M., O.S., A.G., S-Y.L., F.K., and J.W. analysed the data; J.U.H., R.K., I.G.I., and N.T.S. prepared and characterised materials; T.O. contributed to electron beam irradiation; R.N. fabricated solid immersion lenses; M.N. developed software for data acquisition and experimental control; M.N., M.W., and Y-C.C. provided experimental assistance; P.U., J.M., O.S., A.G. and J.W. provided theoretical support; R.N., N.T.S., O.S., A.G., S-Y.L., F.K., and J.W. discussed and wrote the paper. All authors provided helpful comments during the writing process.

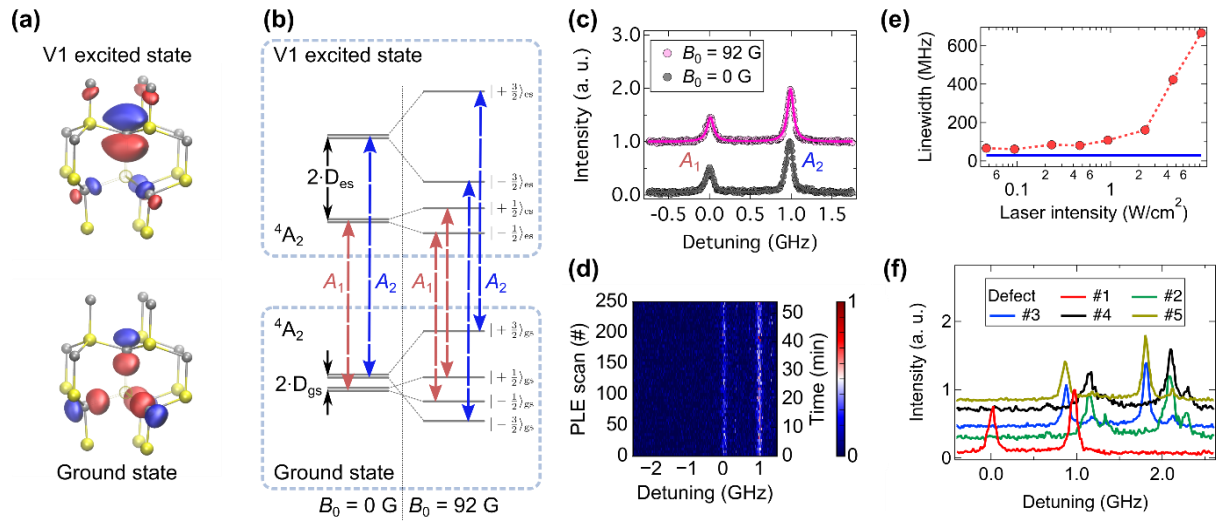

Figure 1: Optical transitions of the silicon vacancy in 4H-SiC. (a) Crystalline structure of 4H-SiC with a silicon vacancy centre at a hexagonal lattice site. Upper (lower) graph shows the square moduli of the defect wave functions of the V1 excited (ground) state, as calculated by density functional theory. Red (blue) shaded areas symbolise that the wave function has a positive (negative) sign. The yellow and grey spheres represent silicon and carbon atoms, respectively, and the crystallographic c-axis is aligned vertically in this figure. (b) Ground and excited state level scheme with and without a magnetic field applied along the c-axis. Red (blue) optical transitions labelled $A_1$ ($A_2$) connect spin levels $m_S = \pm 1/2$ ($m_S = \pm 3/2$). (c) Resonant absorption spectrum of a single vacancy centre at $B_0 = 0$ G and $B_0 = 92$ G. Lines are fits using a Lorentzian function. (d) Repetitive resonant absorption scans at $B_0 = 92$ G over 52 minutes without any sign of line wandering. (e) Absorption linewidth of the peak $A_2$ as a function of the resonant pump laser intensity. Below 1 W/cm² no power broadening is observed and the linewidth is close to transform limited as indicated by the blue line. (f) Resonant absorption spectra of five single defect centres, showing several defects with overlapping lines.

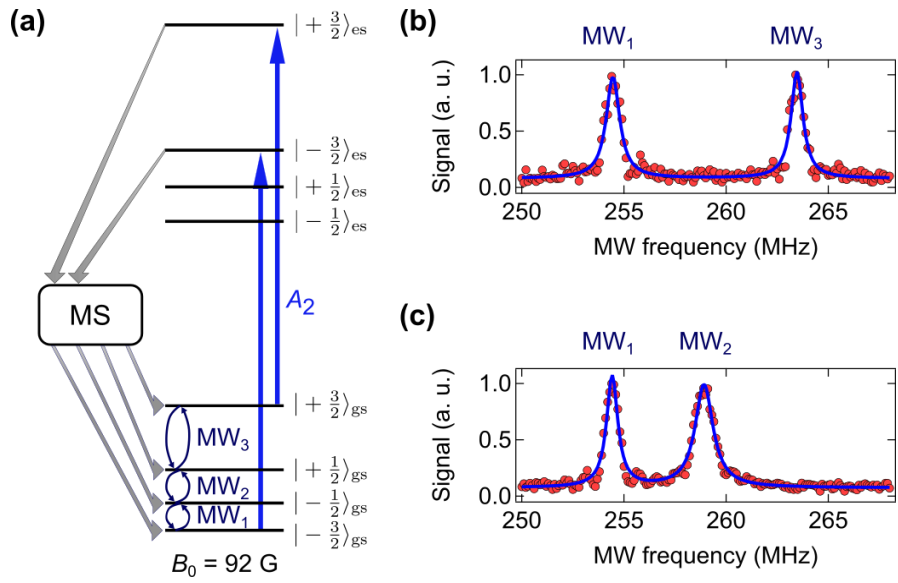

Figure 2: Optically detected magnetic resonance. (a) Level scheme indicating the used optical transition ($A_2$) and microwave fields $MW_1$, $MW_2$ and $MW_3$. Spin flips occur via nonradiative channels involving metastable states (MS). (b) ODMR signal of the ground state after initialising the system into $\left|\pm\tfrac{1}{2}\right\rangle_{gs}$. (c) ODMR signal after initialisation into $\left|-\tfrac{1}{2}\right\rangle_{gs}$. Blue lines are fits using Lorentzian functions. All data are normalised raw data, *i.e.* without background subtraction.

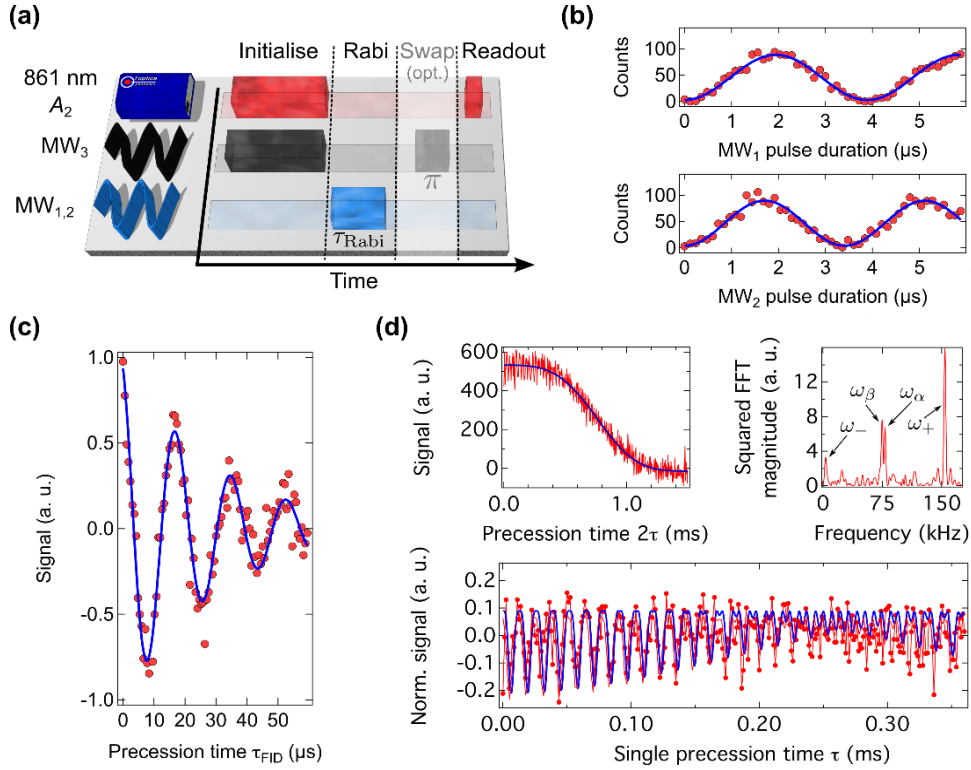

Figure 3: Spin manipulation and coherence. (a) Experimental sequence for observing Rabi oscillations. The system is always initialised into $|-\frac{1}{2}\rangle_{gs}$ using resonant excitation along $A_2$ and MW$_3$. This step is followed by a Rabi sequence (MW$_{1,2}$), an optional population swap ($|+\frac{1}{2}\rangle_{gs} \leftrightarrow |+\frac{3}{2}\rangle_{gs}$), and optical readout. (b) Rabi oscillations for $|-\frac{1}{2}\rangle_{gs} \leftrightarrow |-\frac{3}{2}\rangle_{gs}$ (upper panel) and $|-\frac{1}{2}\rangle_{gs} \leftrightarrow |+\frac{1}{2}\rangle_{gs}$ (lower panel). Blue lines are sinusoidal fits. All data are raw data. (c) Free induction decay measurement yielding $T_2^* = 30 \pm 2$ μs, and the blue line is a fit. (d) Hahn echo measurement and nuclear spin coupling. From the top left graph, we infer $T_2 = 0.85 \pm 0.12$ ms. Red lines are data and the blue line is a fit using a higher-order exponential function. The bottom panel is a zoom into the first part of the Hahn echo after subtraction of the exponential decay function and normalisation. Pronounced oscillations are observed, witnessing coherent coupling to a nearby nuclear spin. Data (red dots connected by lines) are fitted using equation (1) (blue line). The top right panel is a Fourier analysis of the normalised Hahn echo, showing four distinct frequency components through which a weakly coupled $^{29}$Si nuclear spin is identified.

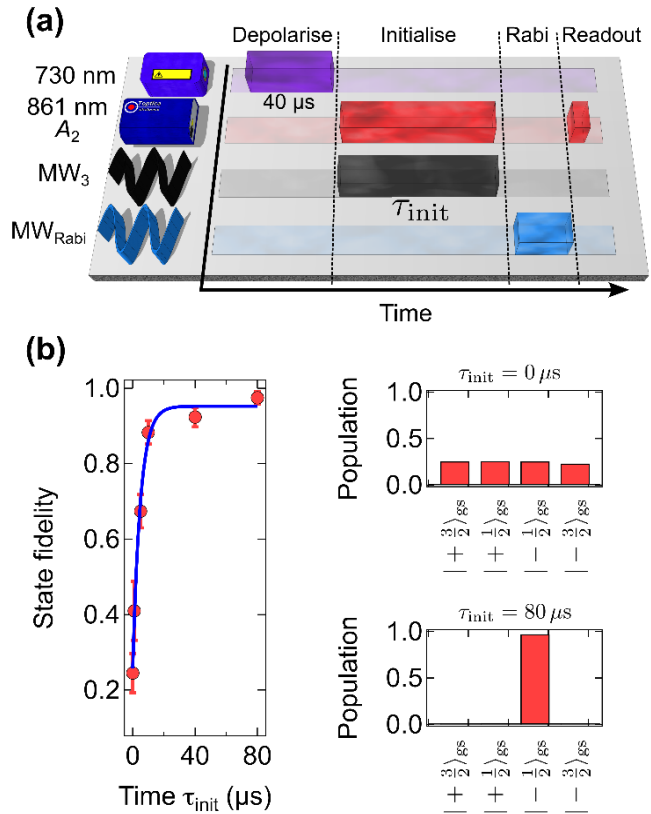

Figure 4: Electron spin initialisation fidelity. (a) Experimental sequence. Before each round, the ground state spin is depolarised using off-resonant excitation for 40 μs. Then, the system is initialised into $|-\frac{1}{2}\rangle_{gs}$. Ground state populations are inferred from Rabi oscillations and resonant optical readout. (b) Left side: Spin population in $|-\frac{1}{2}\rangle_{gs}$ as a function of the duration of the initialisation procedure. Up to 97.5% are achieved. The blue line is a fit using an exponential function. Right side: Inferred spin populations in the four ground state sublevels without initialisation (top) and after 80 μs initialisation time.

**Supplementary information**

**S1. Identification of single silicon vacancy centres.**

To confirm that measurements were performed on single silicon vacancy defects, we performed Hanbury Brown and Twiss (HPT) experiments[1]. To this end, the defect was excited using the off-resonant laser (730 nm) and we recorded the second-order correlation function $g^{(2)}(\tau)$ of the fluorescence emission. A typical result is shown in Figure S1 in which $g^{(2)}(\tau = 0) \ll 0.5$. This stands as a clear proof that a single photon emitter is investigated[2].

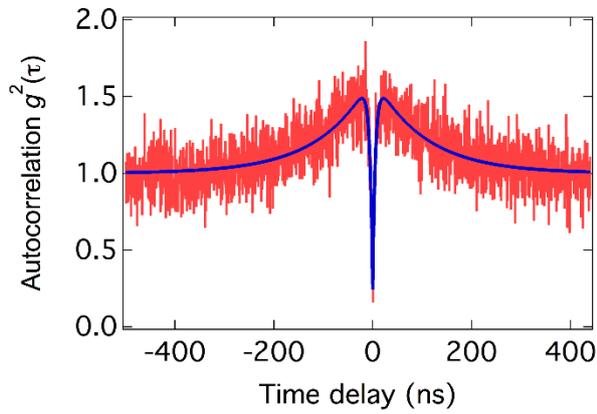

Figure S1: Second-order autocorrelation function. Data are represented in red, and the blue line is a fit assuming a three-level model function: $g^{(2)}(\tau) = \left(1 - \beta \cdot e^{-\frac{|\tau|}{\tau_1}} - (1-\beta) \cdot e^{-\frac{|\tau|}{\tau_2}}\right) \cdot 1/N + (N-1)/N$, in which $g^{(2)}(\tau = 0) = (N-1)/N$, and $N$ is the number of single photon emitting emitters. From this fit, we extract the following parameters: $g^{(2)}(\tau = 0) = 0.24 \pm 0.06$, $\tau_1 = 5.5 \pm 0.4$ ns, and $\tau_2 = 103.7 \pm 3.9$ ns.

## S2. Orientation and polarisation of the optical transitions.

Previous studies have already shown that the silicon vacancy centre at a hexagonal lattice site (V1 centre) is most effectively excited using a linearly polarised off-resonant laser whose polarisation is parallel to the crystal's c-axis[3,4].

Here, we investigate the behaviour of the individual optical transitions $A_1$ and $A_2$ under resonant excitation. As we have shown in the main text, when applying a magnetic field, no additional optical transitions are observed, supporting the absence of circularly polarised optical transitions.

In the following, we show that both transition dipoles are indeed linearly polarised and parallel to each other. To this end, we applied broadband microwaves in order to continuously mix the ground state populations, and performed resonant optical excitation along either $A_1$ or $A_2$ at an intensity of about $1\,\text{W/cm}^2$. The polarisation of the excitation laser was adjusted by a half-wave plate (HWP), the efficiency of the excitation was inferred from the detected fluorescence intensity in the phonon sideband. Figure S2 shows the experimental results. In both cases, tuning the HWP leads to near-perfect sinusoidal oscillations, demonstrating that linear dipole transitions are excited, which are parallel to each other.

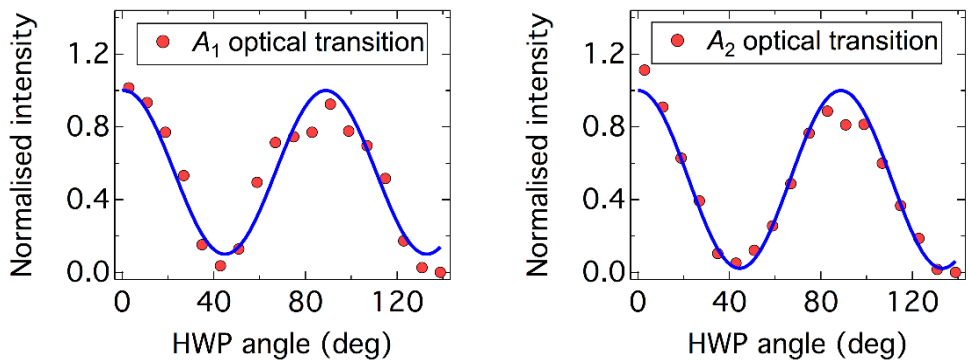

Figure S2: Measuring the polarisation of the optical dipole transitions $A_1$ (left side) and $A_2$ (right side). Dots are data, and blue lines are sinusoidal fits. The high contrast in the observed oscillations shows that both dipoles are linear and parallel to each other.

## S3. Inferring the sign of the excited state zero field splitting.

Using a wavemeter (Coherent WaveMaster) with a resolution of 0.1 GHz, we inferred that the $A_2$ optical transition is the one at higher energy. We now want to infer whether this transition links the spin sublevels $|\pm\frac{3}{2}\rangle_{gs}$ (resulting in a positive sign of $D_{es}$) or the sublevels $|\pm\frac{1}{2}\rangle_{gs}$ (resulting in a negative sign of $D_{es}$). For this, the experimental results of the Rabi oscillations between $|-\frac{1}{2}\rangle_{gs}$ and $|+\frac{1}{2}\rangle_{gs}$ are relevant (see Figure 3(b) in the main text).

We now consider both cases of $D_{es}$ (positive and negative):

Assuming that $D_{es}$ is positive, the $A_2$ optical transition links the $m_S = \pm\frac{3}{2}$ sublevels, such that the ground state spin initialization procedure (laser excitation on the $A_2$ transition and continuous MW$_3$ excitation in order to depopulate $|+\frac{1}{2}\rangle_{gs}$) leads to near deterministic population of $|-\frac{1}{2}\rangle_{gs}$. Thereafter, MW$_2$ is applied for $\tau_{Rabi}$, followed by a population swap ($\pi$-pulse at MW$_3$) between $|+\frac{1}{2}\rangle_{gs}$ and $|+\frac{3}{2}\rangle_{gs}$, and eventual population readout of the states $|\pm\frac{3}{2}\rangle_{gs}$. This experiment should lead to clearly visible Rabi oscillations, starting at a minimum intensity.

Assuming on the other hand that $D_{es}$ is negative, then $A_2$ links the $m_S = \pm\frac{1}{2}$ sublevels, such that the ground state spin initialization procedure would lead to a near deterministic population of $|-\frac{3}{2}\rangle_{gs}$. Thereafter, MW$_2$ is applied for $\tau_{Rabi}$, which does not alter the ground state spin population at all. In addition, the population swap $\pi$-pulse at MW$_3$ has also no effect. As a consequence, the eventual population readout of the states $|\pm\frac{1}{2}\rangle_{gs}$ should lead to no observable signal.

The experimental results in Figure 3(b) in the main text clearly support therefore that the excited state zero field splitting is positive, *i.e.* $2 \cdot D_{es} = 985 \pm 10$ MHz.

We note that our *ab initio* calculations[5] resulted in positive zero-field constant both for the ground and excited states that strongly supports the analysis of experimental results.

**S4. Landé g-factor and optical transition energies of the V1 excited state.**

The spin Hamiltonian of ground and excited states with an external axial magnetic field is:

$$H_{gs,es} = D_{gs,es} \cdot S_z^2 + g_{gs,es}\, \mu_B\, B_0 \cdot S_z.$$

Here, the subscripts gs and es denote ground and excited states, respectively. $2 \cdot D_{gs,es}$ denotes the zero field splitting, $S_z$ is the spin projection operator in the z-direction, $g_{gs,es}$ is the Landé g-factor, $\mu_B$ is the Bohr magneton, and $B_0$ the strength of the axial magnetic field.

Diagonalization of the Hamiltonian leads to four spin-conserving optical transition energies:

$2 \cdot (D_{es} - D_{gs}) + \frac{3}{2} \cdot \mu_B \cdot B_0 \cdot (g_{es} - g_{gs}) + \Delta E_{gs,es}$ for transitions between $m_S = +\frac{3}{2}$ sublevels.

$\frac{1}{2} \cdot \mu_B \cdot B_0 \cdot (g_{es} - g_{gs}) + \Delta E_{gs,es}$ for transitions between $m_S = +\frac{1}{2}$ sublevels.

$-\frac{1}{2} \cdot \mu_B \cdot B_0 \cdot (g_{es} - g_{gs}) + \Delta E_{gs,es}$ for transitions between $m_S = -\frac{1}{2}$ sublevels.

$2 \cdot (D_{es} - D_{gs}) - \frac{3}{2} \cdot \mu_B \cdot B_0 \cdot (g_{es} - g_{gs}) + \Delta E_{gs,es}$ for transitions between $m_S = -\frac{3}{2}$ sublevels.

Here, $\Delta E_{gs,es} \approx 1.44$ eV is the energy difference between ground and excited states. At zero magnetic field, one expects two pairwise degenerate spin-conserving optical transitions for the sublevels $m_S = \pm\frac{1}{2}$ and $m_S = \pm\frac{3}{2}$. For the latter transition ($A_2$ in our notation), we observe a linewidth (full width at half maximum (FWHM)) of $\Delta f_{B_0=0\,G} = 87.6 \pm 1.6$ MHz. At $B_0 \neq 0$ G, one would expect to see four optical spin-conserving transitions, provided that there is a sizeable difference in the Landé factors $g_{gs}$ and $g_{es}$. As shown in Figure 1(c) in the main text, we do not observe any additional transitions at $B_0 = 92$ G. The linewidth of the $A_2$ transition remains essentially unchanged, i.e. $\Delta f_{B_0=92\,G} = 87.7 \pm 1.6$ MHz. Consequently, we conclude that the change in linewidths is $\Delta f = \Delta f_{B_0=92\,G} - \Delta f_{B_0=0\,G} = 0.1 \pm 2.3$ MHz. We assume now that the optical transitions between the sublevels $m_S = \pm\frac{3}{2}$ show a Lorentzian profile, i.e. $I_\pm \propto \dfrac{\left(\frac{a}{2}\right)^2}{\left(f \pm \frac{f_0}{2}\right)^2 + \left(\frac{a}{2}\right)^2}$, in which $I_\pm$ denotes the intensity of the $m_S = \pm\frac{3}{2}$ transition, $a = \Delta f_{B_0=0\,G}$ is the FWHM of each transition at $B_0 = 0$ G, $f$ is the laser frequency offset with respect to the centre of the resonance, and $\pm\frac{f_0}{2}$ is the displacement of the $m_S = \pm\frac{3}{2}$ transition at $B_0 \neq 0$ G. There exists an analytical solution for the apparent FWHM of the sum of two displaced Lorentzian functions:

$$\Delta f_{\text{Double Lorentz}} = \sqrt{2 \cdot f_0^2 - \sqrt{5 \cdot f_0^4 + 8 \cdot f_0^2 \cdot a^2 + 16 \cdot a^4}}.$$

Using this equation, we infer a displacement of $f_0 = 0.2 \pm 11.6$ MHz. By using this result, we now infer the difference in ground and excited state Landé g-factors:

$$g_{\text{es}} - g_{\text{gs}} = \frac{\Delta f_{\text{Double Lorentz}}}{3 \cdot \mu_{\text{B}} \cdot B_0} = (0.5 \pm 30.0) \cdot 10^{-3}.$$

Since previous studies[3] have already reported $g_{\text{gs}} = 2.0028$, we determine the excited state Landé factor to be $g_{\text{es}} = 2.0033 \pm 0.0300$.

In addition, as the ground and excited state *g*-factors have been determined to be nearly identical, if spin-flipping optical transitions ($|\Delta m_S| = 1$) were allowed, they should appear at $\pm g \mu_{\text{B}} B_0 \approx \pm 258$ MHz compared to the spin-conserving transitions. However, such transitions have not been observed as shown in Figure 1(c).

**S5. Ab initio polarisation calculation for the negatively charged silicon vacancy centre in 4H-SiC.**

In the main text, we report studies on resonant optical excitation spectra that show an outstanding spectral stability in contrast to the nitrogen-vacancy (NV) centre in diamond[6]. We attribute the small inhomogeneous distribution (see Figure 1(d) in the main text) to a low sensitivity of the defects to surrounding electric field fluctuations originating from other defects. Since this may be related to a small dipole moment of the V1 centre in 4H-SiC, we test this hypothesis by performing theoretical calculations as described in the following. We calculate the change in the polarisation for the excitation process between $^4A_2$ ground and $^4A_2$ excited states of the negatively charged V1 centre in 4H-SiC (silicon vacancy defect on a hexagonal lattice site). We compare these results with the ones obtained for the nitrogen-vacancy (NV) centre in diamond.

**Computational details**

We apply density functional theory (DFT) for electronic structure calculation and geometry relaxation using the plane-wave-based Vienna Ab initio Simulation Package (VASP)[7–10]. The core electrons are treated in the projector augmented-wave formalism[11]. For the 4H-SiC supercell, calculations are performed with 420 eV plane wave cut-off energy and with $\Gamma$ centred $2 \times 2 \times 2$ k-point mesh to sample the Brillouin zone. For the diamond supercell, we use 420 eV plane wave cut-off energy and $\Gamma$-point to sample the Brillouin zone. We apply Perdew-Burke-Ernzerhof functional in these calculations[12]. The model for the silicon vacancy defect in bulk 4H-SiC is constructed using a 432-atom hexagonal supercell, whereas we use the 512-atom simple cubic supercell to model the NV centre in diamond. The excited state electronic structure and geometry is calculated by constraint occupation of states, or Delta Self-Consistent Field ($\Delta SCF$) method[13].

We calculate the permanent polarisation in ground and excited states, and their difference, in order to infer the coupling to the optical transition. To this end, we use the VASP implementation of both Born effective charge calculation using density functional perturbation theory (DFPT)[14] and the Berry phase theory of polarization[15–17]. In a DFT calculation, one can define the change in macroscopic electronic polarisation ($P$) as an adiabatic change in the Kohn-Sham potential ($V_{KS}$)

$$\frac{\partial P}{\partial \lambda} = \frac{-ife\hbar}{\Omega m_e} \sum_k \sum_{n=1}^{M} \sum_{m=M+1}^{\infty} \frac{\left\langle \psi_{kn}^{(\lambda)} | \vec{p} | \psi_{km}^{(\lambda)} \right\rangle \left\langle \psi_{km}^{(\lambda)} \left| \frac{\partial V_{KS}}{\lambda} \right| \psi_{kn}^{(\lambda)} \right\rangle}{\left( \varepsilon_{kn}^{(\lambda)} - \varepsilon_{km}^{(\lambda)} \right)^2} + c.c.,$$

where $f$ is the occupation number, $e$ the elemental charge, $m_e$ the electron mass, $\Omega$ the cell volume, $M$ the number of occupied bands, $\vec{p}$ the momentum operator, $\lambda$ is the adiabatic parameter, $\varepsilon$ is the band energy. The first part of the equation corresponds to the electronic part of the permanent polarisation ($p_{el}$), whereas the second part corresponds to the contribution of ions ($p_{ion}$) to the permanent polarisation. In a periodic gauge, where the wavefunctions are cell-periodic and periodic in the reciprocal space, the permanent polarisation takes a form similar to the Berry phase expression

$$\Delta P = \frac{ife}{8\pi^3} \sum_{n=1}^{M} \int_{BZ} dk \langle u_{kn} | \nabla_k | u_{kn} \rangle.$$

Using DFPT, $\nabla_k | u_{kn} \rangle$ can be calculated from the Sternheimer equations with similar self-consistent iterations as in DFT:

$$(H_k - \varepsilon_{kn} S_k) \nabla_k | u_{kn} \rangle = \frac{-\partial (H_k - \varepsilon_{kn} S_k)}{\partial k} | u_{kn} \rangle.$$

We determine the radiative transition rate between the ground and excited $^4A_2$ states by calculating the energy dependent dielectric function $\varepsilon_r(E)$. The spontaneous transition rate is given by the Einstein coefficient

$$A = \frac{n\omega^3 |\mu|^2}{3\pi \varepsilon_0 \hbar c^3},$$

where $n$ is the refractive index, $\hbar \omega$ is the transition energy, $\mu$ is the optical transition dipole moment, $\varepsilon_0$ is the vacuum permittivity, and $c$ is the speed of light. $\mu$ is proportional to the integrated imaginary dielectric function ($I$) of the given transition:

$$|\mu|^2 = \frac{\varepsilon_0 V}{\pi} \int \Im \varepsilon_r(E) dE = \frac{\varepsilon_0 V I}{\pi},$$

where $V$ is the volume of the supercell.

**Theoretical results**

The results of the Berry phase evaluation for macroscopic dipole moment calculation are shown in Tables 1 and 2 for the V1 centre in 4H-SiC and the NV centre in diamond, respectively. The change in the total dipole moment is about 20 times larger for NV centre in diamond with respect to that for V1 centre in 4H-SiC. This means that the V1 centre has intrinsically low coupling strength between optical transition and stray electric fields.

Table S1: Macroscopic electric dipole moment of the hexagonal lattice site silicon vacancy defect (V1 centre) as calculated within the Berry phase approximation.

| Transition | $\Delta p_{ion}(e\text{Å})$ | $\Delta p_{el}(e\text{Å})$ | $\Delta p_{tot}(e\text{Å})$ |
|---|---|---|---|
| gr → ex (V1) | 0 | 0.044 | 0.044 |

Table S2: Macroscopic electric dipole moment of the NV centre in diamond as calculated within the Berry phase approximation.

| Transition | $\Delta p_{ion}(e\text{Å})$ | $\Delta p_{el}(e\text{Å})$ | $\Delta p_{tot}(e\text{Å})$ |
|---|---|---|---|
| gr → ex | 0.061 | 0.842 | 0.903 |

**Experimental results and discussion**

Preliminary studies have been performed to constrain $\Delta p_{tot}$ via Stark shift control of optical transition frequencies. To this end, we spanned two parallel copper wires over the sample in order to apply an electric field. The wires were separated by approximately 100 μm and voltages up to $\pm 200$ V were applied. We note that by applying higher electric fields led to electrical breakdown in the cryostats low-vacuum atmosphere. Considering the relative permittivity of 4H-SiC ($\epsilon_r \approx 10$), this results in an estimated in-crystal field of about $\pm 200 \frac{\text{kV}}{\text{m}}$. Two experiments were performed, one in which the electric field was applied along the crystal's c-axis, and a second one in which the field was orthogonal to the c-axis. We performed resonant excitation studies at zero magnetic field ($B_0 = 0$ G) as a function of the electric field strength, in analogy to the studies shown in Figure 1(c). A change in the width or separation of the $A_1$ and $A_2$ optical transitions would indicate electric field sensitivity through which $\Delta p_{tot}$ can be inferred. We observed, however, neither effect within the investigated electric field range. In our current experimental setup, we can infer peak width and separation with a precision of approximately $\pm 12$ MHz. According to theory, an electric field perpendicular to c-axis has no coupling to V1 centre and the coupling is maximum with direction of the electric field parallel to c-axis. By assuming an homogeneous crystal field along the c-axis of $\pm 200 \frac{\text{kV}}{\text{m}}$, we can constrain the Stark shift tuning coefficient to be $\approx 60 \frac{\text{MHz}}{\frac{\text{MV}}{\text{m}}}$. This is approximately two orders of magnitude smaller than reported for NV centres in diamond ($\approx 6.3 \frac{\text{GHz}}{\frac{\text{MV}}{\text{m}}}$)[18], such that we estimate $\Delta p_{tot} < 0.009$. However, theory implies that the coupling coefficient is about an order magnitude smaller for V1 centre in SiC than that for NV centre in diamond. We show below that a compensating field can be developed when both electric field and illumination are applied that can explain the experimental data.

We note that there is a variation in the position of zero-phonon-line (ZPL) for various single V1 centres in the SiC sample of about few 100 MHz but the stability of the ZPL of each investigated single V1 centre is within 60 MHz after repetitive PLE measurements. This implies that charge redistribution occurs upon illumination for each single V1 centre even without applying external electric field. Understanding of these features requires a close inspection of the SiC sample. First of all, the nitrogen donor concentration is very low ($\sim 3.5 \cdot 10^{13}$ cm$^{-3}$), thus the resistivity is very high. This means that the exchange of carriers is very limited, and the concept of Fermi-level does not apply to the entire SiC sample. Si-vacancies are created by 2 MeV irradiation of electrons. According to previous studies[19], the energy threshold of kicking off C and Si atoms in SiC lattice is about 120 keV and 250 keV, respectively. The application of 2 MeV energy should lead to a cascade process where the Si or C atoms kick out other C and Si atoms from the lattice. After annealing at 300 °C, the Frenkel-pairs disappear in the form of either recrystallization or antisite defects. The antisite defects are not activated by near infrared illumination. Besides, carbon interstitial clusters and vacancies may form. Interestingly, Si-vacancy has two faces in SiC crystal: the simple Si-vacancy which is basically a deep acceptor and the carbon antisite-vacancy pair (CAV) complex which has donor levels. Furthermore, the amphoteric carbon vacancies are left after the materials processing. Carbon vacancies are expected to appear in majority because they have the lowest formation energy among the considered defects[5] and threshold energy to form by electron irradiation[19]. The negative charge of the Si-vacancy should be donated by a nearby defect. The average distance between nitrogen donors is about 306 nm in the SiC sample, thus only few Si-vacancies will reside near (~50 nm) the nitrogen donor. Among the considered vacancy-like defects, the CAV defect has the shallowest donor level[5], which can be activated by near infrared illumination. The simplest model is that a positively charged defect resides near the negatively charged Si-vacancy, the V1 centre. According to our calculations, the positively charged donor defect (N-donor or CAV) should sit around ~40 nm around the V1 centre at different lattice sites going from the symmetry axis of the defect toward the basal plane, in order to experience few 100 MHz variation in the ZPL of various V1 centres. The small spectral diffusion may be understood by assuming that another donor defect lies near the V1 centre with about the same distance but another location where the illumination will activate that donor, and the resulting electron, free carrier, will be captured by the previously positively charged donor defect. According to a previous study[20], low energy Si atoms will produce vacancy defects in about 10 nm region; thus Si atoms that are created by 2 MeV electron irradiation have much higher kinetic energy and should produce vacancies, antisites and interstitials at larger distances, around 40 nm and larger distances. By applying an external electric field that is parallel to the symmetry axis of the V1 centre, illumination will again ionize a donor defect but the electric field will drag the electron in the opposite direction of the electric field. One of the carbon vacancies around V1 centre will capture this electron, and the positive donor and negative carbon vacancy will form an electric field that mostly screens the external electric field. We find that if these defects are both 40 nm apart from the V1 centre along the symmetry axis then they shield the external electric field to about 10% of its magnitude. As a consequence, the resulting Stark-shift agrees with the experimental data ($\Delta p_{tot} < 0.09$). Although, this estimation is crude as the statistics about the point defects around V1 centre is not known but our scenario still explains all the experimental findings. We think that both the small coupling constant of the V1 centre and the shielding effects created by the donor and acceptor point defects around the V1 centre are responsible for the spectral stability of the V1 centre.

## S6. Decoherence sources limiting spin coherence times.

In the main text, we report a spin decoherence time of $T_2 = 0.85 \pm 0.12$ ms, measured by Hahn echo. Although this result is better than the previously reported values[4,21–23], one may anticipate reduced decoherence rates owing to the use of an isotopically purified nuclear spin free 4H-SiC sample, in analogy to previous experiments with isotopically purified diamond and silicon[24,25].

As we used a rather low dose of electron beam irradiation to create defect centres, the concentration of paramagnetic defects is small ($\sim 10^{13}$ cm$^{-3}$) as explained in Methods section, which cannot explain the observed $T_2$ times. Shallow nitrogen donors are also discarded as a major decoherence source as their concentration is also too low to be significant ($\sim 3.5 \cdot 10^{13}$ cm$^{-3}$) since the equivalent spin dipole-dipole interaction in electronic spin bath requires higher total impurity concentration ($\sim 6 \cdot 10^{14}$ cm$^{-3}$)[21].

We therefore attribute the main decoherence source to be undesired defects near surface paramagnetic defects created by cutting with a typical concentration in the low $10^{13}$ cm$^{-2}$ range in the region about ~1 μm from the surface[26]. Before irradiation, samples were annealed to 1130 °C in N$_2$ gas flow to reduce the concentration of paramagnetic surface defects to below detection of Electron Paramagnetic Resonance (EPR) (below $10^{12}$ cm$^{-2}$). Since the N$_2$ annealing is known to reduce only 10% of the surface defects and the EPR experimental conditions are not optimized for the detection of the surface defects, it is safe to assume that the surface defect concentration within ~1 μm range from the surface has an upper limit of $10^{16}$ cm$^{-3}$. As the optical transition dipole of the investigated defect centres is parallel to the crystal's c-axis, we had to flip the sample by 90°. This comes with the trade-off that all observed defects are located close to the cutting surface. The cutting surfaces are also expected to have structural defects induced by micro cracks caused by cutting. In the future, this issue can be addressed by improved sample processing[26] or by growing SiC layers on a-plane substrate so that solid immersion lens can be fabricated on as-grown surfaces, which contain no such defects.

**S7. Electronic fine structure and spin polarization of the V1 defect.**

A simplified electronic fine structure model of the V1 defect is shown in Figure S3. The spin $m_S = \pm 1/2$ and $m_S = \pm 3/2$ states of the first excited state (es) are labeled as es$_1$ and es$_2$, respectively. Similarly, the spin states $m_S = \pm 1/2$ and $m_S = \pm 3/2$ belonging to the ground state (gs) are denoted by gs$_1$ and gs$_2$. The optically active transitions es$_1 \to$ gs$_1$ (red transition) and es$_2 \to$ gs$_2$ (blue transition) have equal radiative decay rates as the states involved in these transitions have the same molecular orbital configurations resulting with equal transition dipole moments. Moreover, the spin polarisation of the V1 defect is governed by the phonon-assisted non-radiative intersystem crossing (ISC) from and to the metastable doublet states, ds$_1$ and ds$_2$, via direct spin-orbit coupling (DSO).

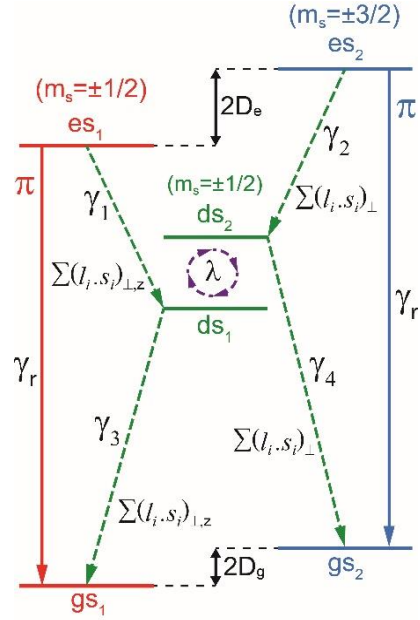

Figure S3: Electronic fine structure of the V1 defect in 4H-SiC.

The four non-radiative ISC decay rates shown in Figure S3 determine the unique features in the PLE spectra of this defect. The non-radiative decay rates in the entry channel that are associated with the decay from es$_1$ to ds$_1$ and from es$_2$ to ds$_2$ are labeled as $\gamma_1$ and $\gamma_2$, respectively, and they determine the amplitudes of the $A_1$ and $A_2$ peaks shown in Figure1(c). Similarly, the decay rates originating from ds$_1$ back to gs$_1$ and from ds$_2$ back to gs$_2$ are labeled as $\gamma_3$ and $\gamma_4$, respectively, and they are responsible for the shelving lifetime of this defect.

We obtain the relationships between $\gamma_1$ and $\gamma_2$, as well as $\gamma_3$ and $\gamma_4$, by evaluating the corresponding DSO matrix elements, $\gamma_i = \frac{2\pi}{\hbar} \left| \langle \psi_i^{gs} | \Sigma_j \, l_j . s_j | \psi_i^{ds} \rangle \right|^2$, in the symmetry adapted $\psi_i^{ds/gs}$ wave-functions basis of $j = 5$ active electrons[27]. We find that 2 metastable doublet states (with symmetries A$_1$ and 2E) have non-zero DSO matrix elements that can participate in the ISC. The E-symmetry ds (e³) can couple to both es and gs by the orthogonal component of the DSO $l_\perp . s_\perp$ (with respect to the c-axis). The remaining A1 ds (ve$^2$) is strongly hybridized with the E ds (e$^3$) by the $l_\parallel . s_\parallel$ component of the DSO. As a result, the overall ISC can be simplified into two doubly degenerate metastable states, labeled as ds$_1$ and ds$_2$, which can only couple to the spin $\pm 1/2$ or the $\pm 3/2$ states of the es and gs, respectively.

The PLE spectra of the V1 defect can be accurately reproduced by using the theoretical fine structure of Figure S3. The PLE signal of the V1 defect corresponds to the sum of the steady state es$_1$ and es$_2$ excited state populations of es$_1$ and es$_2$, that are calculated using the following master equation,

$$\frac{d\rho}{dt} = -\frac{i}{\hbar}[H_0, \rho] + \gamma_r \sum_{i=0}^{2} L(A_r^i) + \sum_{i=0}^{4} \gamma_i L(A_{ds}^i) + \gamma_R \sum_{i=0}^{2} L(A_R) + \gamma_s L(A_s). \quad \text{(eq. S1)}$$

The decay and decoherence processes of the defect are represented by the Lindblad super-operators, i.e. $L(O) = O\rho O^+ - \{O^+ O, \rho\}/2$ for any operator $O$. The electronic fine structure Hamiltonian is given by,

$$H_0 = \frac{(D_{\text{gs}} - D_{\text{es}} + \delta_L)}{2}(|gs_1\rangle\langle gs_1| - |es_1\rangle\langle es_1|) - \frac{(D_{\text{gs}} - D_{\text{es}} - \delta_L)}{2}(|gs_2\rangle\langle gs_2| - |es_2\rangle\langle es_2|) +$$
$$\lambda(|ds_1\rangle\langle ds_2| + |ds_2\rangle\langle ds_1|) + \Omega_L(|gs_1\rangle\langle es_1| + |gs_2\rangle\langle es_2| + c.c.) \quad \text{(eq. S2)}$$

in the rotating frame of the excitation laser with Rabi frequency $\Omega_L$ and detuning $\delta_L = \omega_L - \omega_{\text{ZPL}}$ from the zero phonon line (ZPL). The excited and ground state's ZFS are given by $2D_{\text{es}}$ (975 MHz) and $2D_{\text{gs}}$ (9 MHz), respectively. The radiative decays, $A_r^i = |gs_i\rangle\langle es_i|$, have same radiative decay rate $\gamma_r$. The ISC from es to the metastable ds are given by $A_{ds}^1 = |ds_1\rangle\langle es_1|$ and $A_{ds}^2 = |ds_2\rangle\langle es_2|$ with rates $\gamma_1$ and $\gamma_2$, whereas the ISC from ds back to gs are denoted by $A_{ds}^3 = |gs_1\rangle\langle ds_1|$ and $A_{ds}^4 = |gs_2\rangle\langle ds_2|$ with rates $\gamma_3$ and $\gamma_4$, respectively. The intrinsic spin relaxation in the ground state and dephasing among ds are considered by $A_R = |gs_1\rangle\langle gs_2| + |gs_2\rangle\langle gs_1|$ with rate $\gamma_R$ and $A_s = |ds_1\rangle\langle ds_1| - |ds_2\rangle\langle ds_2|$ with rate $\gamma_s$. Spin-mixing among ds via fast relaxation processes is also included in $H_0$ by the $\lambda$ term.

Resonant optical excitation along the $A_2$ transition (gs$_2 \rightarrow$ es$_2$) will lead to an optical pumping into the gs$_1$. Therefore, to observe a measurable signal during PLE, the ground state spin must be able to relax. Alternatively, one can use MW pulses on the gs spin states for coherent control as well as to overcome this optical pumping. During a continuous broadband MW pulse where all three gs spin transitions shown in Figure 2(a) are allowed to relax equally, the PLE amplitude mismatch between $A_1$ and $A_2$ is directly determined by the ISC rates of $\gamma_1$ and $\gamma_2$. We find that the faster $\gamma_1$ rate causes more population to be removed non-radiatively from $\left|\pm\frac{1}{2}\right\rangle_{\text{es}}$ per optical cycle compared to $\left|\pm\frac{3}{2}\right\rangle_{\text{es}}$, therefore leading to a smaller amplitude for $A_1$ than $A_2$. By comparing the theoretically calculated PLE signal shown in Figure S4 to our experimental results, we deduce that $\gamma_1 \cong 3\gamma_2$.

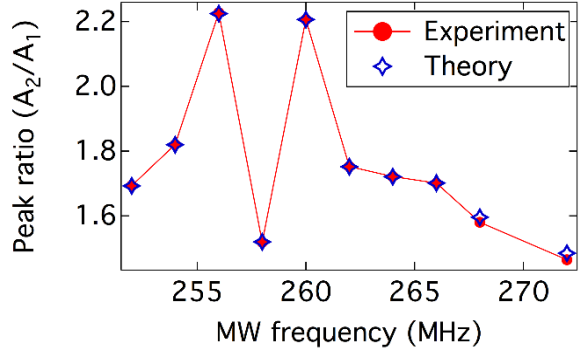

Figure S4: Calculated and measured $A_2/A_1$ ratios for various broadband MW frequencies.

The es ZFS sign, which is determined as in the above section (S3), can also be confirmed by comparing this model to a series of MW schemes in the presence of a magnetic field $B_0 = 96$ G (*i.e.* $g\mu_B B_0 S_z \gg 2D_{gs}$). In these schemes, a continuous broadband MW is scanned from 252MHz to 272MHz and at each frequency the MW centered the $A_2/A_1$ peak ratio is determined from the corresponding PLE spectra. At 258 MHz, all three spin transitions of the ground state (see Figure 2(a)) are allowed to depopulate and the $A_2/A_1$ ratio is solely determined by the $\gamma_1$ and $\gamma_2$ ISC rates. On the other hand, MWs centered at 256 MHz and 260 MHz cannot effectively repopulate the $\left|+\frac{1}{2}\right\rangle_{gs}$ and $\left|-\frac{1}{2}\right\rangle_{gs}$, respectively. This results with increased optical pumping into $\left|+\frac{3}{2}\right\rangle_{gs}$ or $\left|-\frac{3}{2}\right\rangle_{gs}$ state, and causes a significant reduction in the $\left|\pm\frac{1}{2}\right\rangle_{es} \rightarrow \left|\pm\frac{1}{2}\right\rangle_{gs}$ peak amplitude. Therefore, the observed increase in the $A_2/A_1$ ratio for MW frequencies 256 MHz and 260 MHz further supports the positive sign assignment of the es ZFS since it identifies the $A_1$ peak as the $\left|\pm\frac{1}{2}\right\rangle_{es} \rightarrow \left|\pm\frac{1}{2}\right\rangle_{gs}$ transition.